\documentclass{jfm}
\usepackage{soul}
\usepackage{graphicx}
\usepackage{newtxtext}
\usepackage{newtxmath}
\usepackage{natbib}
\usepackage{hyperref}
\hypersetup{
colorlinks = true,
urlcolor = blue,
citecolor = black,
}

\newcommand{\RomanNumeralCaps}[1]
\linenumbers

\title{The minimal flow unit and origin of two-dimensional elasto-inertial turbulence}
\author{Hongna Zhang, Haotian Cheng, Suming Wang, Wenhua Zhang\corresp{\email{zhangwh2022@tju.edu.cn}}, Xiaobin Li, \and Fengchen Li\corresp{\email{lifch@tju.edu.cn}}}

\affiliation{State Key Laboratory of Engine, Tianjin University, Tianjin 300350, PR China}

\begin{document}
\maketitle

\begin{abstract}
The research on elasto-inertial turbulence (EIT), a new type of turbulent flow, has reached the stage of identifying the minimal flow unit (MFU). On this issue, direct numerical simulations (DNSs) of FENE-P fluid flow in two-dimensional channels with variable sizes are conducted in this study. We demonstrate the existence of MFU for EIT to be self-sustained. When the channel length is relatively small, a steady arrowhead (SAR) structure with a laminar-like friction coefficient is observed at high Weissenberg number ($Wi$). However, as the streamwise length increases, the flow fully develops into EIT, characterized by high flow drag. We think the flow falling back to “laminar flow” is caused by the insufficient channel size below the MFU. Furthermore, we give the relationship between the MFU and the effective $Wi$ and explain its physical reasons. By capturing the onset and development process of EIT benefiting from MFU, we confirm that EIT originates from the wall mode rather than the center mode. Moreover, the fracture and regeneration of polymer extension structures observed is the key mechanism for the self-sustaining of EIT.
\end{abstract}

\begin{keywords}
Minimal flow unit, elastio-inertial turbulence, viscoelastic fluid, direct numerical simulation
\end{keywords}


\section{Introduction}
\label{sec:intro}

Viscoelastic fluids widely exist in nature and the unique rheological properties give rise to different flow behaviors comparing with Newtonian fluids, such as drag-reducing turbulence (DRT) at a moderate or high Reynolds number (Re) (\cite{Li12}) and elastic turbulence (ET) at an extremely low Re (\cite{Steinberg21}). The recent discovery of a new type of turbulent state (elasto-inertial turbulence, EIT) by \cite{Samanta13} opened up new avenues for viscoelastic turbulence. Unlike Newtonian inertial turbulence (IT) and ET, EIT arises from the combined effects of nonlinear elasticity and fluid inertia. Its distinctive characteristics are trains of spanwise cylindrical vortex structures of alternating sign around sheets of high polymer extension (\cite{Dubief13}, \cite{Choueiri18}, \cite{Shekar19}, \cite{Dubief23}). The understanding of EIT has provided valuable insights into the maximum drag reduction (MDR) state of DRT as well as the so-called early turbulence of viscoelastic fluids. Recently, \cite{Morozov22} discovered the presence of two-dimensional traveling wave solutions of EIT in channel flow and the induced arrowhead structure, providing a strong link between ET and EIT.

The explorations of the origin and the self-sustaining mechanisms of EIT are hot topics, with focuses on whether it originates from wall modes or center modes as well as whether the transition occurs in a subcritical or supercritical manner (\cite{Dubief23}). Here, the wall mode and center mode are characterized with phase speed close to the critical-layer velocity near channel wall and the maximum base-flow velocity at channel center, respectively. On one hand, through linear stability analysis, \cite{Garg18} firstly discovered the linear instability of center mode under high $Wi$, which is considered as the origin of EIT (\cite{Page20}, \cite{Khalid21}). Weakly nonlinear analysis by \cite{Wan21} indicate that the transition to EIT is subcritical at low polymer concentrations and supercritical at high polymer concentrations. Numerical simulations revealed that the subcritical nonlinear evolution of the center mode can induce saturated "arrowhead" traveling waves (\cite{Page20}), and the steady arrowhead (SAR) structure identified by \cite{Dubief22} and \cite{Beneitez23} is considered as a signature of the center mode, supporting the idea of linear instability. In the experiments of viscoelastic pipe flow (\cite{Choueiri18}, similar arrowhead structures of center mode were observed at low Re, suggesting the significance of center mode instability in the origin of EIT. On the other hand, Graham group proposed the critical layer theory and the nonlinear routine to EIT induced by Tollmien-Schlichting (TS) wall-mode (Shekar et al., 2019, 2020, 2021). As the linear stability analysis predicted the viscoelastic fluid flow stable under all the numerical conditions, they argued that the nonlinear amplification of the viscoelastic fluid wall mode (TS mode) termed as viscoelastic nonlinear TS attractor (VNTSA), is the origin of EIT. 

So far, significant progress has been made on the origin and the mechanism of EIT, but the picture is still incomplete and numerous questions remain open. Direct numerical simulation (DNSs) can shed crucial light on these questions, which however faces the challenges of well-known high Weissenberg number problem (HWNP) and lack of a numerical criterion. Although great efforts have been made in dealing with HWNP over the past two decades, it has not been completely solved (\cite{Alves21}). Most of the numerical studies on EIT adopts some artificial diffusion to alleviate the hyperbolicity of the constitutive model at high $Wi$. However, the hyperbolicity is not only critical to the existence of EIT but also necessary for its sustainability. In addition, unlike the numerical criterion that is well-documented for Newtonian IT, it has not been established for EIT yet, such as how large is the domain required to excite and sustain EIT. The improper choice of computational domain sometimes leads to different picture of dominant dynamics. For example, as found in \cite{Dubief22} the simulated flow states are related with the computational domain, an increase in the streamwise length can change the SAR state to the EIT state under the same parameters. \cite{Zhu21} found slightly higher friction factors compared to the original size results after enlarging the computational domain, which they explained as spatial intermittency and correlation on longer length scales. In our previous studies on EIT of Oldroyd-B fluids (\cite{Zhang22}), we found that a longer channel is required to excite EIT as $Wi$ increases. Therefore, a standard criterion for DNS of EIT after solving HWNP is a prior and plays a crucial role in drawing a complete picture of EIT dynamics.

To establish this criterion, it is worth emphasizing the importance of finding the minimal flow unit (MFU) that can ensure the occurrence and continuous self-sustenance of turbulence (\cite{Jiménez91}). In numerical simulations of turbulence, it is commonly required to ensure that the length of the computational domain is sufficiently long to allow for adequate turbulence development and that the streamwise correlation reaches zero within half of the domain length (\cite{Dubief23}). The statistical results based on the MFU agree with those obtained from significantly larger flow units. The use of MFU provides reliable guidance for selecting the computational domain size in numerical simulations, ensuring that the results contain complete and accurate physical information while reducing the computational cost associated with large domains. It has played an important role in understanding the self-sustaining mechanisms of turbulent structures therein and has remained widely employed in studies related to Newtonian wall-bounded turbulence (\cite{Yin18}). However, few studies are conducted on the MFU of viscoelastic turbulence, particularly EIT. \cite{Xi10} explored the MFU for drag-reducing turbulence in three-dimensional channel flow. Later,  \cite{Graham14}  mentioned the Newtonian fluid MFU cannot sustain turbulence at high Wi. Like Newtonian IT, the MFU is also a necessity to further promote the understanding of EIT including SSP of coherent structures, exact coherent structures (ECSs) and so on. 

This paper aims at finding the size of MFU that is able to sustain EIT, thereby discussing its origin. To this end, a comprehensive investigation on the computational domain effects on the flow characteristics in a wide range of parameters is required based on reliable numerical methods. In our recent work, we identified the improper interpolation of the tensor field when solving the constitutive equations as the main cause of HWNP (\cite{Zhang23}). Instead of component-based interpolation, we proposed a tensor-based interpolation method for the conformation tensor and have demonstrated the effectiveness of the tensor-based interpolation method in resolving the challenges of HWNP with no need for artificial diffusion term. This efficient and stable numerical method offers us the ability to access the numerical criterion of EIT today. Moreover, the two-dimensional (2D) nature of EIT demonstrated by \cite{Sid18} implies the MFU of EIT in a channel is mainly determined by the streamwise length of the computational domain. Therefore, we conduct a series of numerical simulations on 2D plane Poiseuille flow to find the MFU suitable for EIT and thereby explore its origin. The remaining sections are organized as follows: Section 2 introduces the governing equations of viscoelastic fluid flow, numerical methods and conditions; Section 3 discusses the computational domain effects on flow characteristics and explores the origin of EIT based on its MFU; Section 4 gives the conclusions.

\section {Numerical methodology}

This study focuses on the 2D plane Poiseuille flows of FENE-P fluid under constant flow rate. Channel walls are assumed to be non-slip, and the periodic boundary condition is applied in the streamwise direction. Taking the channel half height $h$, the volume-averaged velocity $u_b$ ($u_b = \frac{1}{2h}\int_0^{2h} U(y)dy $ with $U(y)$ the locally averaged velocity in the streamwise direction), $h/u_b$, and $\rho u_b^2$ as the reference length, velocity, time and pressure, respectively, the dimensionless governing equations of FENE-P fluid flow in the form of conformation tensor \textbf{c} are as follows:
\begin{equation}
\nabla \textbf{u} = 0\label{EQ.1},
\end{equation}

\begin{equation}
\frac{\partial \textbf{u}}{{\partial t }} + \textbf{u} \cdot \nabla \textbf{u} = - {\nabla p}+ \frac{\beta}{{{\mathop{\rm Re}\nolimits} }} {\nabla^2 \textbf{u}} + \nabla \cdot \boldsymbol{\tau} \label{EQ.2},
\end{equation}

\begin{equation}
\boldsymbol{\tau} =\frac{{1-\beta}}{{\rm {Re Wi}}}[ f (r)\textbf{c} - \textbf {I}]\label{EQ.3},
\end{equation}

\begin{equation}
\frac{\partial \textbf{c}}{{\partial t }}+(\textbf{u} \cdot \nabla)\textbf{c}- \textbf{c} \cdot (\nabla \textbf{u})- (\nabla \textbf{u})^{\rm T} \cdot \textbf{c}=-\frac{f(r)\textbf{c} - \textbf {I}}{\rm Wi}\label{EQ.4},
\end{equation}
where, \textbf{u} is the velocity vector with $(u, v)$ denoting the streamwise $x$ and wall-normal $y$ velocity components; \textbf{c} is the conformation tensor representing the average of the end-to-end vector of the polymer molecules taken over all the molecules; $p$ is the pressure; $\tau$ is the additional elastic stress tensor; $\beta = \eta_s/\eta_0$, and $\eta_0$ is the solution dynamic viscosity and $\eta_s$ is the solvent contribution to the viscosity; $Re = \rho u_bh/\eta_s$ is the bulk mean Reynolds number; $Wi = \lambda u_b/h$ is the Weissenberg number based on the relaxation time $\lambda$ of the viscoelastic fluid; $f(r) = (L^2 - 3)/(L^2 - r^2)$ is the Peterlin function with $r^2 (r^2 = \rm{tr(\textbf{c}}))$ the trace of the conformation tensor {\bf c}.

The governing equations are solved based on the finite difference method using a DNS code developed in our previous work. Detailed numerical procedures and validation can be found in Zhang et al. (2022, 2023). The tensor-based interpolation method is adopted to deal with HWNP. The key of this method is the interpolation of the eigenvalues and orientation of the conformation tensor, which is more physically motivated. Compared to traditional component-based interpolation method, the accuracy of the conformation tensor's invariants as well as the SPD property of the conformation tensor can be guaranteed at high $Wi$. Additionally, it can be combined with high-order numerical schemes, such as high-order total variation diminishing schemes, to further improve the numerical accuracy. The application procedures are as follows: (i) decomposing the conformation tensor field {\bf c} as 
\begin{equation}
\rm \textbf{c} = \textbf{R} \rm \boldsymbol{\Lambda} \textbf{R}\label{EQ.5},
\end{equation}
\begin{equation}
\rm \boldsymbol{\Lambda} = \begin{bmatrix}
\lambda_1 & 0 & 0 \\
0 & \lambda_2 & 0 \\
0 & 0 & \lambda_3 \\
\end{bmatrix}\label{EQ.6},
\end{equation}
\begin{equation}
\textbf{R} = \begin{bmatrix}
\cos\theta \cos\varphi & \sin\psi \sin\theta \cos\varphi-\cos\psi \sin\varphi & \cos\psi \sin\theta \cos\varphi+\sin\psi \sin\varphi \\
\cos\theta \cos\varphi & \sin\psi \sin\theta \sin\varphi+\cos\psi \cos\varphi & \cos\psi \sin\theta \sin\varphi-\sin\psi \cos\varphi \\
-\sin\theta & \sin\psi \cos\theta & \cos\psi \cos\theta \\
\end{bmatrix} \label{EQ.7},
\end{equation}
where, $\psi$, $\theta$ and $\varphi$ are Euler angles relative to the Cartesian coordinate system; (ii) given the known conformation tensor field $\textbf{c}$, obtain the rotation matrix $\textbf{R}$ and the diagonal matrix $\boldsymbol{\Lambda}$ by Eq. \ref{EQ.5}, and calculate the Eulerian angles and eigenvalues at the grid nodes; (iii) obtain the Eulerian angles ($\psi_{i+1/2}$, $\theta_{i+1/2}$, $\varphi_{i+1/2}$) and eigenvalues ($\lambda_{1, i+1/2}$, $\lambda_{2,i+1/2}$, $\lambda_{3,i+1/2}$) at the grid interface through various interpolation schemes; (iv) calculate the diagonal matrix $\boldsymbol{\Lambda}_{i+1/2}$ and rotation matrix $\textbf{R}_{i+1/2}$ at the grid interface by Eqs. \ref{EQ.6} and \ref{EQ.7}; (v) reconstruct the conformation tensor $\textbf{c}_{i+1/2}$ at grid interface by Eq. \ref{EQ.5}. 

A series of simulations of EIT state are conducted in a 2D channel with varying dimensionless channel length $SX$, where $SX = 5n$ (with $n$ = 0.5, 1.0, 1.6, 2.0 and 4.0, respectively).  In the existing studies ({\it e.g.,} Dubief et al., 2013, 2022), a dimensionless channel length $SX$ of 5 is frequently used, but no discussion on whether it can satisfy the MFU criterion. A wide range of $Wi$ is covered from 10 to 800, while keeping $Re = 2000, \beta = 0.9$ and $L^2 = 10000$. Linear stability analysis indicates that center mode instability occurs when $Wi > 70$ (\cite{Cheng23}). However, despite the linear stability analysis predicting linear stability for $Wi$ between 10 and 70 (with the center mode being the least stable in this parameter range), DNSs conducted by Shekar et al. (2020, 2021) demonstrate that EIT can indeed be excited at $Wi \textgreater 10$. Here, due to the use of different characteristic velocities to define Wi, the case of $Wi=10$ in our study corresponds to $Wi =15$ in \cite{Shekar20}. During the simulation, the grid resolution is set to be $256n\times304$, and the time step size is chosen to be $5\times10^{-4}h/u_b$ or even smaller. Each simulation runs for a duration of at least $2500h/u_b$ to achieve statistical convergence.

\section {Results and analysis}
\subsection {Computational domain effects on flow states}

Firstly, the effects of the computational domain on flow states are evaluated through the statistical property, specifically the flow drag. Figure 1 illustrates the ensemble-averaged drag coefficient ($C_f$) of viscoelastic flows over a wide range of $Wi$ obtained by different channel lengths. The two inset figures display the temporal evolution of the spatially-averaged drag coefficient at two representative $Wi$ of 40 and 100. It is evident that the choice of the computational domain plays a crucial role in determining the numerically achieved flow states, particularly for cases of large $Wi$ ({\it{e.g.}}, above 40). When longer channels ({\it{e.g.}}, $SX=10$ and 20) are employed, the temporal evolution of $C_f$ and flow structures demonstrate continuous occurrence of EIT, reaching a self-sustained state at $Wi \textgreater 10$. In these cases, the statistical drag coefficients converge as the channel length increases from $SX=10$ to 20. This indicates that a computation domain with $SX \textgreater 10$ is sufficient to capture the EIT state of FENE-P fluid for all presently considered $Wi$ at $Re = 2000$, $\beta=0.9$ and $L^2 = 10000$. Furthermore, it is observed that the converged $C_f$ exhibits a power-law increase with Wi within these two channels. Interestingly, the power-law exponent is approximately 0.17, which closely resembles the value of about 0.2 found for ET in channel flow (\cite{Steinberg21}). This implies that the elastic effect on EIT shares similarities with ET. Taking into account of the nonlinear extension effect of the FENE-P model, the scaling of $C_f$ with elasticity becomes simpler. Here, $Wi_r$ is defined to characterize the average effective elastic effect for the FENE-P fluid as $Wi_r=\int_0^2 \overline{{\frac{Wi}{f(r)} \frac{\partial u}{\partial y}}}dy$. In terms of $Wi_r$, the scaling simplifies to a beautiful linear one: $C_f \propto Wi_r$. It indicates that $Wi_r$ is more suitable to describe the elastic effect in EIT, and the drag of EIT linearly depends on the effective elastic effect. 

For the shorter channel lengths, such as $SX=5$ and 8, the numerical results align with those of the longer channels at lower Wi ({\it{e.g.}}, $Wi \textless 40$ for $SX=5$ and $Wi \textless 60$ for $SX=8$). However, as $Wi$ is further increased, the flow alternates between an active turbulent state with fluctuating high drag and a hibernating state with stable low drag across a wide range of parameters. Consequently, the statistical $C_f$ decreases significantly, approaching levels seen in laminar flow ({\it{e.g.}} $Wi\textgreater60$). In Figure 1, red open circles and closed circles distinguish the drag coefficients of these two states at the same $Wi$. Notably, the averaged $C_f$ in the high-drag state follows the scaling of EIT obtained from the longer channels, while the averaged $C_f$ in the low-drag state approaches laminar flow. Through evaluating the detailed flow field, it is discovered that the hibernating low-drag state in these cases corresponds to the SAR structure state identified by \cite{Dubief22} as illustrated in Figure 3. Moreover, in the cases with $SX=8$, both states exist intermittently for a certain duration (as shown in Inset 2 of Figure 1), while for $SX=5$, the high-drag state immediately transitions to the low-drag state with increase of $Wi$. This suggests that the high-drag state cannot continuously sustain in a short channel, and a long channel is required to capture the EIT state. It is worth noting that the intermittent flow state observed in the case with $SX=8$ bears resemblance to the intermittent maintenance of laminar and turbulent states reported by \cite{Shekar21}. The influence of the computational domain on flow states mimics the effects of $Wi$. Here, the channel length of $SX=8$ is close to the critical length or the MFU to excite a continuous EIT state at $Wi \textgreater 60$. These findings can also be used to explain the effects of $L^2$ obtained by \cite{Dubief22}. Therein, they discovered that drag increase (DI) rises with increasing $Wi$ under small $L^2$ conditions, then reaches a maximum value and subsequently decreases to zero (laminar flow) under large $L^2$ conditions. It is argued that the above phenomenon is caused by the insufficient channel length used for large $L^2$.

\begin{figure}
\centering
\includegraphics[width=0.55\textwidth]{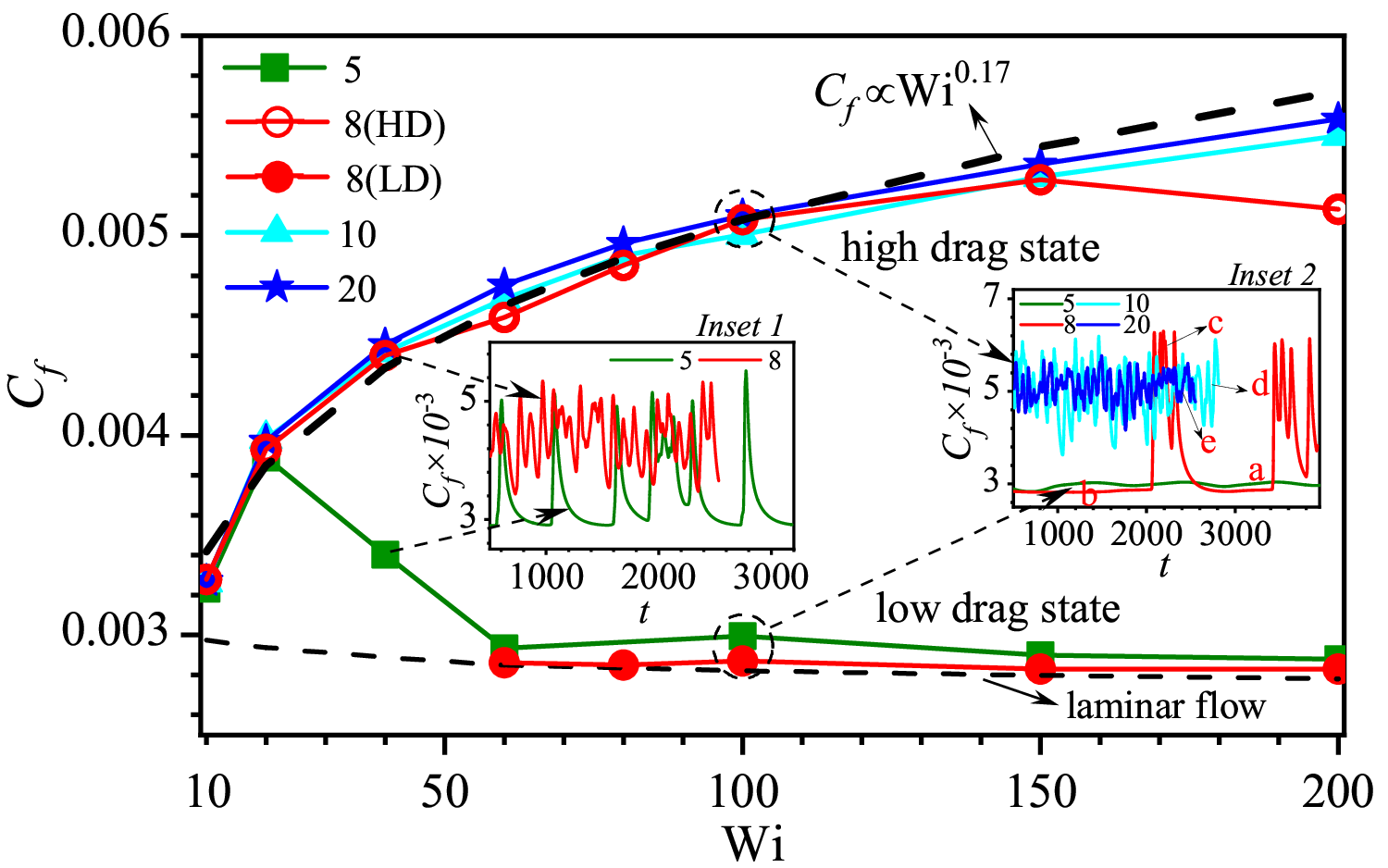}
\caption{\label{fig01} Statistical $C_f$ of viscoelastic fluid flow at different $Wi$ obtained by different computational domain. Two insets are the instantaneous $C_f$ at two typical $Wi$ of 40 and 100, respectively. Note: The red open and closed circle symbols represent the average $C_f$ for different stable stages at $SX=8$ under the same $Wi$. The dashed line corresponds to $C_f$ in laminar regime of FENE-P fluid at different $Wi$.}
\end{figure}

According to the obtained drag coefficients and flow fields, Figure 2 presents a phase diagram summarizing various viscoelastic flow states obtained from different computational domain sizes in a wide range of $Wi$ ( $Wi \textgreater 10$). The flow states depicted in the diagram are stable ones that can be sustained for extended periods. As aforementioned, the cases at all $Wi$ considered in this paper are capable of reaching EIT. This diagram allows us to identify the MFU required to produce EIT state for different $Wi$. It is evident from the diagram that the size of the MFU increases with $Wi$ and saturates when $Wi$ exceeds a critical value ({\it e.g.}, 100). Given the case of Wi above the critical value to excite EIT, if the computational domain is slightly shorter than the size of MFU, a coexistence of EIT and SAR appears. Otherwise, EIT cannot be excited and only SAR structures appear if the computational domain is too short. For example, when $Wi <40$, a flow unit with a size of $SX\textgreater 5$ can achieve self-sustained EIT state. However, for $Wi \textgreater  40$, a flow unit with $SX=5$ is no longer able to sustain the EIT state and instead exhibits a coexistence of SAR and EIT, or even a pure SAR state. Further increasing $Wi$ above 60, a flow unit with $SX=8$ also loses its ability to sustain the EIT state and shows a coexistence of SAR and EIT. Notably, the flow unit with $SX=10$ is sufficient to sustain EIT states for $Wi$ ranging from 10 to 200 or even larger Wi, indicating a saturation of the effective elastic effect of FENE-P fluid at high $Wi$. After supplementing a significant amount of numerical database, we give the size of MFU required to sustain the EIT state in numerical simulations as $SX \textgreater f(Wi)$ for the parameters investigated in this study. This criterion can also be expressed as $SX \textgreater{(0.65Wi_r -10.66)}$ in terms of $Wi_r$. Appropriate size of the flow unit satisfying this criterion is suggested for the numerical simulation of EIT.

\begin{figure}
\centering
\includegraphics[width=0.55\textwidth]{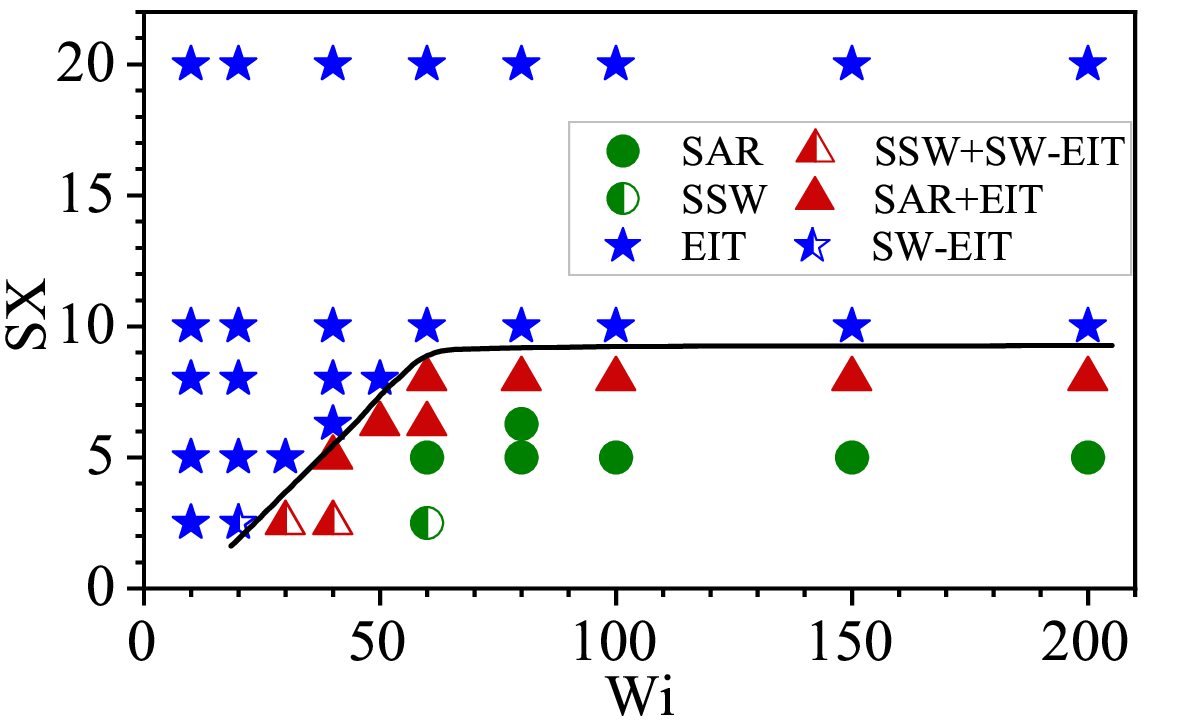}
\caption{\label{fig2} Phase diagram of flow states at different $Wi$ obtained by different computational domain. SSW: steady sheets near the wall; SW-EIT: EIT occurrence near single wall; "+": the alternative occurrence of different states.}
\end{figure}

The above results clearly demonstrate that the existence of MFU for EIT to be self-sustained. This naturally raises the following questions: Why long channel is required to produce EIT? What determines the MFU of EIT in numerical simulations? Answering these questions can give more insights into the origin and self-sustaining mechanism of EIT. Through evaluating the flow fields, it is observed the evolution process of sheet-like structures of polymer extension plays a crucial role in the generation and self-sustaining of EIT (See Supplementary Movies), which is in line with the findings in \cite{Shekar21}. Therefore, we postulate that proper channel length to capture the evolution process of these structures is essential to sustain the EIT state. To test this hypothesis, we focus on the numerical results of cases obtained from a sufficiently large computational domain of $SX=20$. Figure 3(a) illustrates the spectral characteristics of the elastic energy at different $Wi$. A peak in the low wavenumber can be observed near the wall at various $Wi$, corresponding to the characteristic length of the sheet-like structures or the spacing between two sheets. Additionally, Figure 3(b) summarizes the relationship between the characteristic size of the sheets and $Wi$, which exhibits a consistent pattern with the critical MFU determined in Figure 2. Based on these observations, we argue that the size of the MFU required for EIT is determined by the characteristic scale of the sheet-like structures. These findings can be used to establish numerical criterion for EIT and theoretical studies based on the concept of the MFU.

\begin{figure}
\centering
\includegraphics[width=0.8\textwidth]{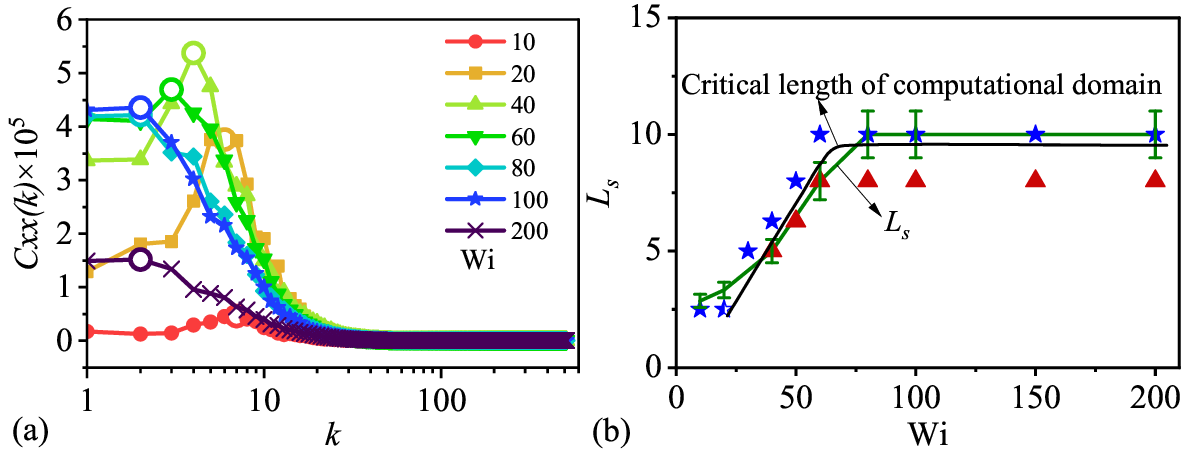}
\caption{\label{fig3} (a) Spectra of the streamwise elastic energy $C_{xx}$ at $y^*\approx 0.5$ at different $Wi$; (b) Characteristic scales $L_s$ of the $C_{xx}$ and the critical computational domain at different $Wi$, where $L_s = 20/k_{p}$ and $k_{p}$ is the peak wavenumber obtained from $C_{xx}(k)$. }
\end{figure}

\subsection {Intermittent flow regime}

This part explores the origin and self-sustaining mechanism of EIT based on the identification of the MFU. Figure 4 draws the dynamical and structural evolution of typical flow cases. In Figure 4a, $C_f$ and $\tau_{xx}$ characterize the energy supply and turbulent intensity of channel flow. For the cases at $SX=20$, it can be seen from the phase diagram that there are elliptical envelopes during the dynamical evolution of EIT, and the principal axes of the elliptical envelopes almost overlap under these conditions, which implies that the selected physical quantity can well describe the dynamics of EIT. For the channels with $SX=8$, the envelope of dynamical evolution also has overlapping principal axes under low $Wi$ conditions ($Wi \textless 40$). However, at high $Wi$, the flow shuttles back and forth between SAR and EIT. Starting from SAR, at first, $C_f$ only shows a small increase, while  $\tau_{xx}$ is significantly enhanced; subsequently, $C_f$ shows a rapid increase and falls back to the EIT envelope line, while  $\tau_{xx}$ is no longer significantly increased; finally, the flow falls back to SAR from the envelope line along the principal axis and repeats the above process. Figures 4b and 4c focus on the channel flows at $Wi=100$. As $SX$ increases, the flow undergoes the evolution process of pure SAR, SAR-EIT, and fully developed EIT, respectively. The flow envelope almost coincides under the conditions of $SX=10$ and $SX=20$. The above phenomenon fully demonstrates that the flow entering the center mode is caused by insufficient channel length, and EIT does not originate from the center mode. 

Furthermore, the flow process of SAR-EIT switching provides a fabulous opportunity for exploring the origin and self-sustaining mechanism of EIT, as shown in Figure 4c ($SX=8, Wi=100$). In addition, we provide movies of the dynamical and structural evolution process, as shown in the supplementary materials with Case A ($SX=5, Wi=40$) and Case B ($SX=8, Wi=100$). Firstly, a near-wall sheet-like streamwise extension structures at $y^*\approx0.5$ appear in the SAR flow regime (see state 1), which gradually grow (see states 2 and 3) and begin to split (see the movies of Case B). Subsequently, frequent splitting (see dynamic figure) leads to a rapid increase in $C_f$ and the flow develop into EIT (see state 4). Finally, EIT could not be maintained and the near-wall extension structures gradually decline (see state 5), and the flow enters the stable SAR regime dominated by the center mode (see state 6). It should be emphasized that the initially excited near-wall extension structures are not induced by the SAR. Thus. it can be drawn that EIT originates from the wall mode rather than the center mode. The self-sustaining process of EIT can be presented from the perspective of coherent structure regeneration: small disturbances in the flow dominated by the wall mode induce high extension sheet-like structures, which gradually grow and split under the elastic nonlinearity as well as the fluid inertia, regenerating new turbulent coherent structures. The above process continues to occur, maintaining the turbulent state of the flow. Thus, there exists a structural similarity in the EIT: large extension sheets generate small extension sheets, and small extension sheets continues to regenerate as they grow. This is different from the process of large eddies generating small eddies and small eddies generating mini eddies in inertial turbulence.

\begin{figure}
\centering
\includegraphics[width=1\textwidth]{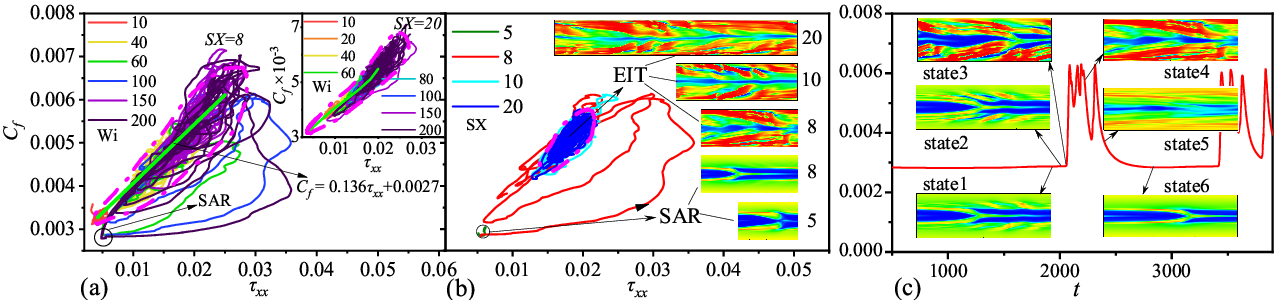}
\caption{\label{fig4} Dynamical and structural evolution cycles of (a)  $C_f$ with $\tau_{xx}$ for different Wi at $SX = 8$ and $SX = 20$; (b) $C_f$ with $\tau_{xx}$ for different $SX$ at $Wi =100$ (inset figures present the snapshots of streamwise extension fields $C_{xx}$ obtained by different $SX$ corresponding to the typical state); (c) structure intermittency at $Wi = 100$ and $SX = 8$.}
\end{figure}

\begin{figure}
\centering
\includegraphics[width=1\textwidth]{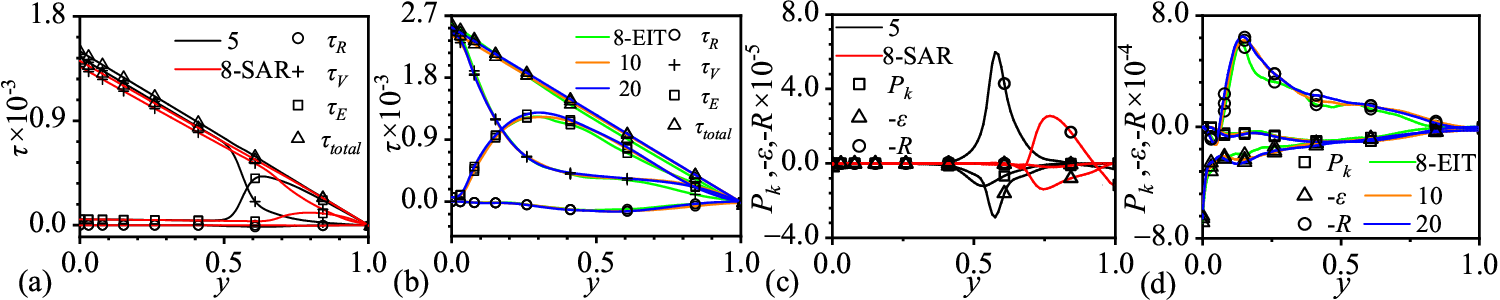}
\caption{\label{fig5} Statistical properties for different periods: (a) Stress and (b) Velocity profile in the normal-wise direction based on inner scale; (c) and (d) turbulent kinetic energy budget in the normal-wise direction for different channels. 8-SAR and 8-EIT correspond to the results obtained by $SX=8$ in the SAR and the EIT stage, respectively.}
\end{figure}

The above mentioned origin of EIT can be further confirmed from the perspective of potential flow dynamics as shown in Figure 5. The SAR induced by the center mode exhibits significant centralization characteristics: (1) The elastic nonlinear shear stress is tightly supported near the channel center with the peak close to the center (see Figure 5a); Although additional nonlinear shear stress is formed relative to laminar flow, it cannot cause a significant increase in flow drag due to the low shear strain rate therein; (2) Compared to negative $P_k$, turbulent kinetic energy comes from energy conversion $R$, which is also tightly supported near the channel center (see Figure 5d). However, EIT exhibits significant near-wall characteristics: (1) the peak of elastic nonlinear shear stress locates much closer to the walls compared to the SAR state (see Figure 5b), which can induce high flow drag weighted by high shear strain rate there; (2) The formation of turbulent kinetic energy also relies mainly on energy conversion $R$, but the peak of $R$ is very close to the wall. The above phenomena indicate that there exit completely different dynamic mechanisms for EIT from the center mode. Observing the time evolution process of the above physical quantities in the supplementary material (see the movies of Case A), it can be found that the peak of elastic nonlinear shear stress caused by the near-wall extension structures inducing EIT is close to the wall, and so is the energy conversion R. $\tau_{E}$ and R gradually develop to the shape and magnitude at EIT condition. However, the dynamic process corresponding to the center mode under $SX=8$ gradually declines with the occurrence of EIT. These phenomena once again prove the above arguments on the origin of EIT.

\section {Concluding remarks}

In summary, a series of DNSs of 2D channel FENE-P fluid flow are performed to obtain the MFU of EIT in this paper. Based on the MFU, the origin and self-sustaining process of EIT are then investigated. Major conclusions can be drawn as follows. The numerical results show that if the channel length is long enough, the flow will sustain the EIT regime with the increase of Wi, otherwise, the flow will gradually fall back from the EIT state to the SAR state due to the insufficient channel length. This implies the presence of MFU in the EIT. We found that the MFU is essentially determined by the characteristic scale of polymer extension structures whose evolution is crucial to the sustenance of EIT. In the absence of SAR structures, the sheet-like extension structures near the wall at $y^{\ast}\approx 0.5$ can be induced and gradually evolve into fully developed EIT. This means that EIT does not originate from the center mode, but from the wall mode which induces the sheet-like extension structures near the wall. In detail, the sheet-like extension structures induced by small disturbances gradually grows and fractures, while the extension structures formed by fractures continue to grow and fracture. Once triggered, EIT is self-sustained due to the regeneration of the extension structures. So far, the above results are obtained under fixed inertial effect, and comprehensive investigations are of course still necessary to draw an exhaustive picture of MFU for EIT. Moreover, with MFU, now we can reach the level of identification of various ECS, the geometry of EIT and detailed dynamical process of EIT in the future work.

\section*{Acknowledgements}
This research was funded by the National Natural Science Foundation of China (NSFC 51976238, 52006249, 12202308).

\section*{Declaration of interests}
The authors report no conflict of interest.


\bibliographystyle{jfm}


\end{document}